# A node-wise analysis of the uterine muscle networks for pregnancy monitoring

N. Nader, M. Hassan, W. Falou, C. Marque and M. Khalil

*Abstract*— The recent past years have seen a noticeable increase of interest in the correlation analysis of electrohysterographic (EHG) signals in the perspective of improving the pregnancy monitoring. Here we propose a new approach based on the functional connectivity between multichannel (4x4 matrix) EHG signals recorded from the women's abdomen. The proposed pipeline includes i) the computation of the statistical couplings between the multichannel EHG signals, ii) the characterization of the connectivity matrices, computed by using the imaginary part of the coherence, based on the graph-theory analysis and iii) the use of these measures for pregnancy monitoring. The method was evaluated on a dataset of EHGs, in order to track the correlation between EHGs collected by each electrode of the matrix (called 'node-wise' analysis) and follow their evolution along weeks before labor. Results showed that the strength of each node significantly increases from pregnancy to labor. Electrodes located on the median vertical axis of the uterus seemed to be the more discriminant. We speculate that the network-based analysis can be a very promising tool to improve pregnancy monitoring.

## I. INTRODUCTION

The interest in connectivity has increased recently for a variety of bio-signals such as EEG [1] ECG [2] and electrohysterogram (EHG) [3], [4]. The EHG is the noninvasive abdominal measurement of the uterine electrical activity [5]. It has been considered as promising for clinical applications in obstetrics, such as monitoring of pregnancy, labor detection, and prediction of preterm labor [5].

Numerous studies have shown the importance of analyzing the propagation of the uterine electrical activity for the discrimination between pregnancy and labor contractions. It is also been used for pregnancy monitoring, computed as conduction velocity (CV) [7], [8] or propagation velocity (PV), together with nonlinear correlation coefficient [9]. PV combined with peak frequency (PF) showed so far the highest performance to discriminate labor and nonlabor contractions [6]. Recently, it has been shown that the nonlinear correlation between EHG leads increases from pregnancy to labor [8]. A comparative study, performed between several correlation measures applied to EHG signals [9], showed high variability among the tested methods and evidenced EHG propagation to the whole matrix and in all directions [9].

Using another approach, Escalona-Vargas et al. used the magnetomyography (MMG) in order to understand the mechanism of activation of the uterine muscle. They study with MMG, the propagation velocity of the uterine activity during pregnancy [10]. They found that the propagation was multidirectional and ranged from 1.9-3.9 cm/s.

In most of previous studies, the EHG correlation matrices were always reduced by keeping only their mean and standard deviation. Despite the encouraging results obtained, relevant information was missed due to this averaging. To characterize precisely the correlation matrix and quantify the associated connectivity, the graph theory analysis seems to be a more pertinent tool. According to this approach, a correlation matrix can be represented as a graph consisting of a set of nodes (electrodes) interconnected by edges (connectivity values between electrodes) [3], [4].

In a very recent study [3], we showed that the combination of Imaginary part of coherence, used as connectivity method, and of the Strength of the obtained graph, Strength(Icoh), showed the highest performance to classify pregnancy and labor contraction, when compared to the most efficient parameters reported until now (Conduction Velocity [7] and the propagation velocity combined with peak frequency method (PV+PF) [6]).

In this paper, we show the pertinence of the proposed approach in monitoring pregnancy. The originality of this work consists in the node-wise analysis that focuses the analysis of the graph parameters to each node separately.

## II. MATERIALS AND METHODS

### A. Overview

The complete pipeline of our approach is presented Fig. 1. The first step consists of recording the uterine contractions by using a grid of 4*4 electrodes (Fig. 1a). The EHG signals are then segmented and denoised (Fig. 1b). The third step is to compute the connectivity between the denoised signals by using different connectivity methods (Fig. 1c). The obtained connectivity matrix is then represented by a graph (Fig. 1d). These graphs are then computed from uterine pregnancy and labor EHG bursts at different term (Fig. 1e). Several measures can be extracted from the obtained graphs based on graph theory (Fig. 1f). These measures are used to evaluate the clinical pertinence of this approach for the classification between labor and pregnancy contractions (Fig. 1g).

*Research supported by Lebanese University and University of Technology of Compiegne.

N. Nader is with Azm Center in Biotechnology, Lebanese University, Tripoli, Lebanon and with Sorbonne University, Université de technologie de Compiègne, CNRS, UMR 7338 BMBI, 60200 Compiègne, France (Phone: +33 7 55 81 57 55; +961 70 436 909; e-mail: noujoud.nader@utc.fr).

M. Hassan is with LTSI, Université de Rennes 1, F-35000, France (mahmoud.hassan@univ-rennes1.fr).

W. Falou and M. Khalil are with Azm Center in Biotechnology and its application, Lebanese University, Tripoli, Lebanon (wafalou99@hotmail.com, Mohamad.khalil@ul.edu.lb ).

C. Marque is with Sorbonne University, Université de technologie de Compiègne, CNRS, UMR 7338 BMBI, 60200 Compiègne, France (Catherine.marque@utc.fr).

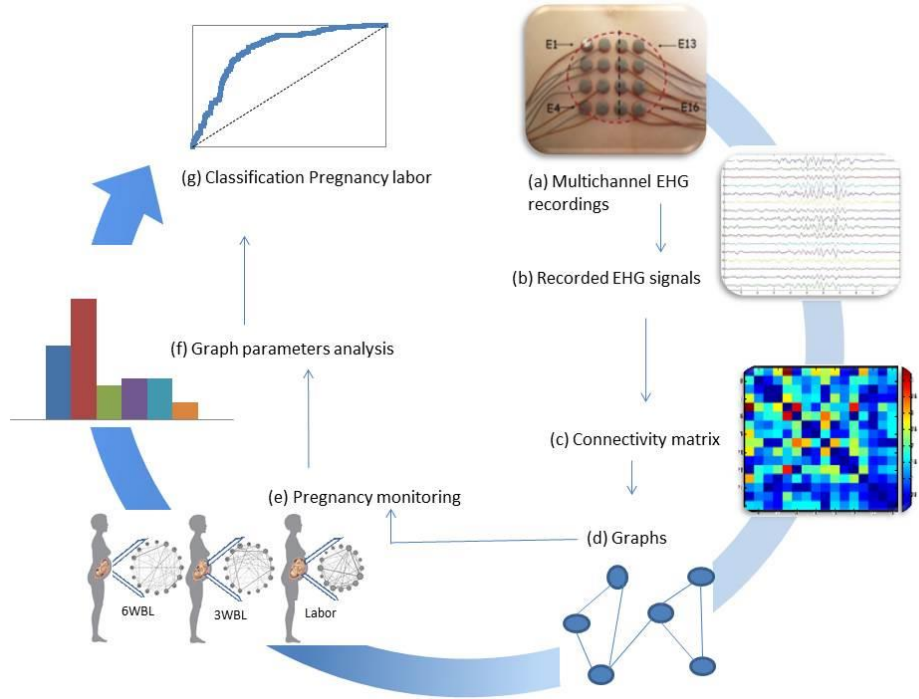

Figure 1. Structure of the investigation. (a) multichannel EHG recordings using a grid of 4*4 electrodes. (b) Recorded segmented and filtered EHG signals. (c) Evaluted Connectivity matrix using connectivity methods. (d) Graph Computation (e) Graphs used for pregnancy monitoring along week of gestation . (f) Statistical study based on the extraction of graph parameters. (g) Classification labor/pregnancy using graph metrics.

*B. Data*

The signals were recorded from 16 monopolar channels of a 4x4 matrix located on the woman's abdomen. The third electrode column was always put on the uterine median vertical axis (see [11] for more details). The sampling frequency was 200 Hz. The data were recorded at the Landspitali university hospital (Iceland) using a protocol agreed by the relevant ethical committee (VSN02-0006-V2) and at the Center for Obstetrics and Gynecology (Amiens, France), using a protocol agreed by the relevant ethical committee (ID-RCB 2011-A00500-41). The EHG signals were manually segmented and denoised by using a CCA-EMD method developed in our team [12]. After segmentation and denoising, we obtained 183 labor and 247 pregnancy EHG bursts at different weeks of gestation. The analysis below has been applied to these segmented uterine bursts. The obtained pregnancy and labor EHG bursts were grouped in 11 groups of week before labor (WBL) and a labor group. We have presented here only the WBL groups that contain more than 4 women and more than 20 contractions per group. Thus we keep the following groups: 8WBL, 6WBL, 4WBL, 3 WBL, 2 WBL, 1WBL and Labor.

*C. The Imaginary part of Coherence (Icoh)*

The coherence (C) function gives the linear correlation between two signals X and Y as a function of frequency. It is defined as their cross-spectral density function $C_{XY}$ normalized by their individual auto-spectral density functions $C_{XX}$ and $C_{YY}$. Nolte et al. [13] have showed that the use of the imaginary part of the coherence function can reduce the effect of the conducting volume (abdominal muscle, fat and skin layers present between the uterus and the abdominal electrodes). In this paper, we evaluate the imaginary coherence (Icoh) defined as:

$$\text{Icoh} = \frac{|\text{Im}C_{XY}(f)|}{\sqrt{|C_{XX}(f)||C_{YY}(f)|}} \quad (1)$$

*D. Graph Theory*

Once the connectivity matrix is calculated, from the Icoh, it is then transformed into a graph. A graph is an abstract representation of a network, consisting of a set of nodes (N) and edges (V), indicating the presence of an interaction between the nodes [14]. An analysis based on the graph theory consists in extracting different parameters from the graph. Here we computed the strength of each node which shows the importance and contribution of each node with respect to the rest of the network.

$$S_i = \sum_{j \in N} w_{ij} . \quad (2)$$

where i, j denotes respectively the $i^{th}$, $j^{th}$ nodes, and $w_{ij}$ is the value (Icoh value) of the relation between nodes i and j [1].

## III. RESULTS

*A. Classification Labor vs Pregnancy*

The *Icoh* method was applied to each burst of the labor and pregnancy databases. The matrices were thresholded to keep the values above the mean of the matrix in order to remove the insignificant coupling. Each matrix was then transformed into a graph, from which we extracted the strength parameter. We obtain thus for each contraction the corresponding strength value for each electrode. Finally, we obtain for each electrode (node) 247 strength values for pregnancy and 183 strength values for labor. To investigate the possible difference between pregnancy and labor, we plotted a boxplot for these two classes on each electrode. Results are shown Fig. 2. The figure shows an increase in the strength values from pregnancy to labor with noticeable differences for all the electrodes. We used the Wilcoxon test in order to test the significance of these differences. The results, shown Table 1, indicate that all the differences between labor and pregnancy Strength(Icoh) are significant if $p<0.01$, corrected for multiple comparison using Bonferroni method, whatever the electrode. Thus, if we compare two groups (labor and nonlabor) within sixteen subsets (16 electrodes) the treatments will be significantly different at the 0.01 level if there is a P value less than 0.01/16 within any of the subsets.

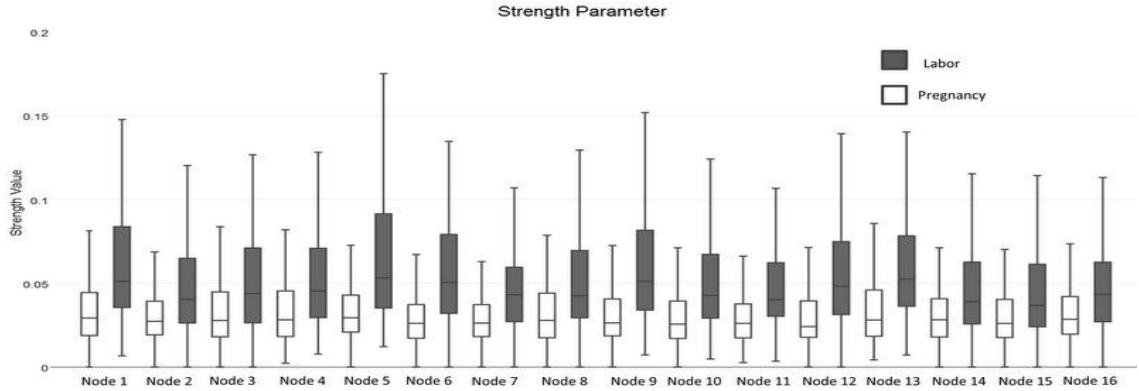

Figure 2. Boxplot of strength parameter values in pregnancy and labor on 16 nodes (electrodes). All the differences are significant ($p<0.01$).

### B. Pregnancy monitoring on node 12.

To evaluate, for a given electrode, the evolution of Strength(Icoh) along term, we have computed the value of Strength(Icoh) for each node and each available week of gestation group. A typical example is showed Fig. 3 for node number 12.

Fig. 3a shows that all the strength values during pregnancy stay relatively low. We cannot see clear differences between the term groups from 8WBL to 1WBL, while an increase between 1WBL and Labor groups is noticeable.

In order to evidence more the evolution of the node 12 connectivity, we present in Fig. 3b-h the corresponding averaged graphs for each of the term groups. We highlighted in each graph only node 12 and the nodes to which it connects. The thickness of the edge represents the weight (Icoh value) and the diameter of a node represents its strength. We can notice in the Labor graph that node 12 is associated to a high number of significant edges (Fig. 5h). It is indeed connected to 11 nodes over 16, unlike during pregnancy, where node 12 connects always to a maximum of 6 nodes, whatever the pregnancy group (Fig. 3b-g). In terms of node strength (diameter of the node), no clear difference can be noticed between all the WBL graphs. However, during labor node 12 is clearly larger (higher strength) than for the other pregnancy groups.

We then computed the statistical differences between all the terms by using the Wilcoxon test. No significant difference was observed between the pregnancy groups ($p>0.01$), except between 8WBL and 2WBL ($p=0.009$). A significant difference was always obtained between Labor and all the other groups ($p<0.01$).

### IV. DISCUSSION

We have used in this work a grid of 4*4 electrodes for the recording the uterine EHG signals and computed the connectivity between these signals. We then studied the strength of each node (electrode). First, we tested the ability of this parameter for the classification of pregnancy and labor contractions. The highest classification rate was observed at node 5, 1 and 12 (node 12 is positioned on the median vertical axis of the woman's abdomen [15] ).

Second, we have presented the ability of the proposed approach to improve pregnancy monitoring. We thus parted the contractions into 6 groups of weeks before labor and a labor group. A slight difference was observed in Strength(Icoh) between the gestation groups, while a noticeable evolution was observed for the Labor group, indicating an increase in connectivity from pregnancy to labor. This is in agreement with the many previous results mainly those using the nonlinear correlation coefficient method [8]. A possible explanation for this increase in connectivity during labor is the propagation phenomenon, associated with the presence of gap junctions that increase in number just before and during labor [16]. Another hypothesis could be the presence of the hydrodynamic-stretch activation mechanism, recently made by Young [17] to explain the uterine synchronization noticed during labor. We are aware that more effort and experience need to be realized to confirm one of these explanations.

TABLE I. WILCOXON TEST RESULTS BETWEEN LABOR AND PREGNANCY AT EACH NODE (ELECTRODES)

| Nodes | P_value | Nodes | P_value |
|---|---|---|---|
| Node 1 | 5.76E-11 | Node 9 | 6.94E-11 |
| Node 2 | 1.33E-07 | Node 10 | 1.72E-08 |
| Node 3 | 6.53E-05 | Node 11 | 8.69E-10 |
| Node 4 | 7.14E-08 | Node 12 | 2.79E-10 |
| Node 5 | 4.93E-13 | Node 13 | 8.66E-08 |
| Node 6 | 1.15E-11 | Node 14 | 0.00111 |
| Node 7 | 8.71E-09 | Node 15 | 0.000119 |
| Node 8 | 7.69E-07 | Node 16 | 0.000322 |

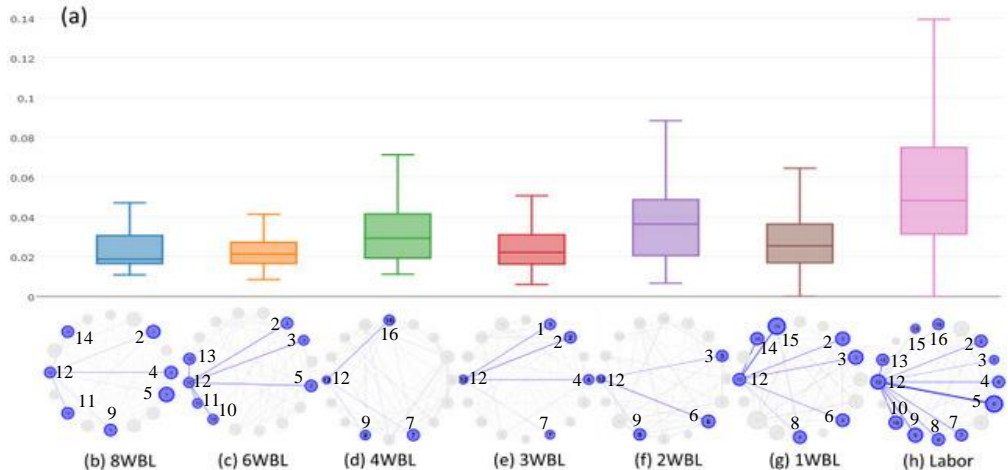

Fig 3. (a) Boxplot of strength vales for node 12 from with week before labor. Mean graph for: (b) 8WBL. (c) 6WBL. (d) 4WBL. (e) 3WBL. (f) 2WBL. (g) 1WBL. (h) Labor.

## V. CONCLUSION

We reported here a node-wise analysis of EHG correlations during pregnancy and labor. Strength(Icoh) values showed a significant increase from pregnancy to labor which mean that a high correlation is happening between different parts of uterus during labor. Electrodes at the median axis of the uterus seemed to be the more discriminant nodes. We think that this network-based approach is a very promising tool to quantify uterine synchronization for a better pregnancy monitoring.

We expect this approach to be further useful for a better classification between pregnancy and labor and thus would help for the early prediction of preterm labor.